\let\KernelLabel\label
\documentclass[aps,prl,reprint,amsmath,amssymb,superscriptaddress,nobibnotes,nofootinbib,longbibliography,floatfix]{revtex4-2}
\usepackage[T1]{fontenc}
\usepackage[utf8]{inputenc}
\usepackage{graphicx}
\usepackage{xcolor}
\usepackage{microtype}
\makeatletter
\@ifl@t@r\fmtversion{2023-06-01}{%
  \let\label\KernelLabel
  \let\ltx@label\label
}{}
\makeatother
\usepackage[hypertexnames=false,colorlinks,citecolor=blue,urlcolor=blue,linkcolor=blue,filecolor=blue]{hyperref}
\makeatletter
\AtBeginDocument{\immediate\write\@auxout{\string\citation{apsrevTitledBibliography}}}
\makeatother
\newcommand{\SIsection}[2]{S#2}
\AddToHook{cmd/normalsize/after}{%
  \setlength{\abovedisplayskip}{8pt plus 2pt minus 4pt}%
  \setlength{\belowdisplayskip}{8pt plus 2pt minus 4pt}%
  \setlength{\abovedisplayshortskip}{8pt plus 2pt minus 4pt}%
  \setlength{\belowdisplayshortskip}{8pt plus 2pt minus 4pt}%
}
\begin{document}
\title{Microscopic Mechanism and Control of Reentrant Localization}
\author{Pei-Jie Chang}
    \thanks{These authors contributed equally to this work.}
    \affiliation{State Key Laboratory of Low-Dimensional Quantum Physics and Department of Physics, Tsinghua University, Beijing 100084, China}

\author{Quanfeng Lu}
    \thanks{These authors contributed equally to this work.}
    \affiliation{State Key Laboratory of Low-Dimensional Quantum Physics and Department of Physics, Tsinghua University, Beijing 100084, China}
    \affiliation{Beijing Academy of Quantum Information Sciences, Beijing 100193, China}

\author{Wentao Yang}
    \thanks{These authors contributed equally to this work.}
    \affiliation{State Key Laboratory of Low-Dimensional Quantum Physics and Department of Physics, Tsinghua University, Beijing 100084, China}

\author{Dong Ruan}
    \affiliation{State Key Laboratory of Low-Dimensional Quantum Physics and Department of Physics, Tsinghua University, Beijing 100084, China}
    \affiliation{Frontier Science Center for Quantum Information, Tsinghua University, Beijing 100084, China}

\author{Gui-Lu Long}
    \email{gllong@tsinghua.edu.cn}
    \affiliation{State Key Laboratory of Low-Dimensional Quantum Physics and Department of Physics, Tsinghua University, Beijing 100084, China}
    \affiliation{Beijing Academy of Quantum Information Sciences, Beijing 100193, China}
    \affiliation{Frontier Science Center for Quantum Information, Tsinghua University, Beijing 100084, China}
    \affiliation{Beijing National Research Center for Information Science and Technology, Beijing 100084, China}
\begin{abstract}
Increasing quasiperiodic modulation can restore extended states after localization, raising the question of how the same modulation drives both localization and recovery. We identify a microscopic mechanism in quasiperiodic dimer chains by resolving scattering between Fourier channels. Specifically, scattering into neighboring channels and back shifts the spectral edge relative to a localization boundary determined by a neutral auxiliary resonance. Their successive crossings drive the same spectral packet with fixed integrated-density-of-states (IDS) labels into localization and back into an extended regime. This recovery persists across finite parameter ranges in both gap-opening families, including relative-phase detuning. Furthermore, an exact sublattice mapping enables compensated control of the recovery window and predicts recovery thresholds for individual IDS labels.
\end{abstract}
\maketitle

Anderson localization shows that disorder can suppress quantum transport~\cite{anderson1958,evers2008}. Localization can also arise in quasiperiodic lattices without randomness~\cite{sokoloff1985}. For example, the Aubry--Andr\'e--Harper model has a single modulation threshold~\cite{aubry1980,jitomirskaya1999}; generalized onsite and hopping modulations allow mobility edges and coexistence of extended and localized spectra~\cite{dassarma1988,biddle2009,biddle2010,ganeshan2015,li2017,wang2020mosaic}. Transport therefore depends on which states remain extended and on their fraction of the spectrum~\cite{purkayastha2017,roy2021fraction}. Atomic and photonic platforms probe these distinctions through expansion and state-resolved measurements~\cite{roati2008,lahini2009,luschen2018,an2021mobility,vaidya2023}.

Yet the same modulation that localizes can, beyond a further threshold, restore extended states. Roy et al. established reentrance in a hopping-dimerized chain with staggered quasiperiodic modulation~\cite{roy2021}. Related recovery phenomena occur in other quasiperiodic lattices~\cite{padhan2022,wei2025} and Aubry--Andr\'e--Fibonacci interpolations~\cite{goblot2020,tabanelli2024}. Further work examines critical behavior~\cite{roy2022critical} and longer-range hopping effects~\cite{wang2023longrange,chang2025longrange,allard2026}. Experiments have also observed reentrant localization in photonic quasicrystals~\cite{vaidya2023} and photonic SSH lattices with random-dimer disorder~\cite{xu2024observation}. Complementary duality-based approaches locate mobility edges and phase boundaries in quasiperiodic chains~\cite{goncalves2022hidden,goncalves2023critical}. Our earlier work identified multiple reentrant transitions in an onsite-staggered chain with a composite potential~\cite{chang2026}. This onsite-staggered chain and Roy et al.'s hopping-dimerized chain provide distinct gap-opening routes. Numerical studies of both models emphasized competition between the two-site structure and quasiperiodic modulation. For these dimer chains, however, the remaining questions are why recovery windows open and close, which spectral states recover, and how their recovery can be controlled.

We connect a microscopic explanation of reentrance to an exact control protocol. Fourier-channel scattering shifts the central spectral edge relative to a zero-growth branch set by a neutral auxiliary resonance. Their crossings produce capture and release of a common IDS packet. We map recovery regions as hopping dimerization, onsite staggering, and relative phase vary. An exact mapping between equal and unequal sublattice modulations provides a compensation path that shifts release at fixed recapture and locates recovery thresholds for selected IDS labels.

\begin{figure}[!t]
\centering
\includegraphics[width=83mm,trim=0 .2mm 0 .2mm,clip]{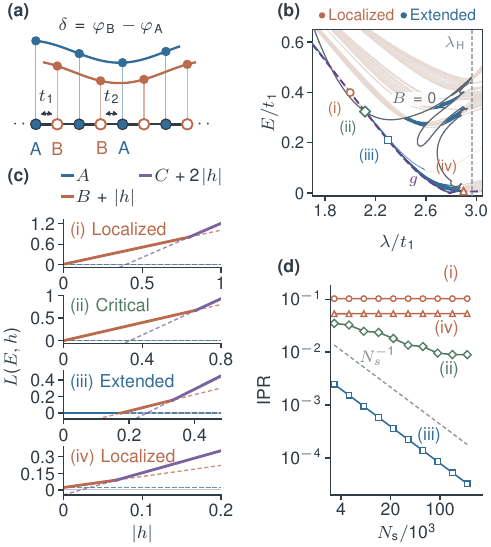}
\caption{Reentrance at the central-gap edge. (a) Dimer chain with hoppings $t_1,t_2$ and relative phase $\delta$. (b) Gray: 3194-site spectrum; blue/orange: growth-based extended/localized samples. Gray curves: zero-growth contour $B=0$; purple dashed: edge $g$; vertical dashed: recovery ceiling $\lambda_H=2\sqrt{t_1t_2}$. Markers (i--iv): $\lambda/t_1=2.00,\lambda_c/t_1,2.30,2.90$, with $\lambda_c/t_1\simeq2.12$ (critical release). (c) Three-branch LE envelopes; dashed: individual branches. (d) IPR scaling over ten sizes $3194\leq N_s\leq242786$; dashed: $N_s^{-1}$. Methods: SM Sec.~\SIsection{sec:sm-fig12}{4}.}
\label{fig:abc-growth}
\end{figure}

\emph{Model and three-branch growth law.---}\phantomsection\label{sec:model}
We consider a quasiperiodic dimer chain with sites $A$ and $B$ in each unit cell and intracell and intercell hoppings $t_1,t_2>0$ [Fig.~\ref{fig:abc-growth}(a)]. Its Hamiltonian is given by
\begin{equation}
\begin{aligned}
H={}&\sum_m\bigl(t_1c_{m,A}^{\dagger}c_{m,B}
 +t_2c_{m+1,A}^{\dagger}c_{m,B}+\mathrm{H.c.}\bigr)\\
 &+\sum_{m,s}u_s(\theta_m)c_{m,s}^{\dagger}c_{m,s},
\end{aligned}\label{eq:hamiltonian}
\end{equation}
where $c_{m,s}^{\dagger}$ ($c_{m,s}$) creates (annihilates) a particle on sublattice $s=A,B$ in unit cell $m$. The onsite potential combines a sublattice offset $v_s$ with a quasiperiodic modulation,
\begin{equation}
u_s(\theta_m)=v_s+\lambda\cos(\theta_m+\varphi_s),\quad
\theta_m=\theta_0+2\pi m\omega ,\label{eq:onsite}
\end{equation}
with amplitude $\lambda$ and Diophantine frequency $\omega$. Here $\theta_0$ is the initial phase, and the sublattice phase offsets $\varphi_s$ define the relative phase $\delta=\varphi_B-\varphi_A$ modulo $2\pi$. Energies and modulation strengths are measured in units of $t_1$. The hopping model has $t_1\neq t_2$ and $v_A=v_B=0$, opening the central gap through hopping dimerization. By contrast, the onsite model has $t_1=t_2$ and $(v_A,v_B)=(-V,V)$, with the gap opened by onsite staggering.

The reference examples use $t_2/t_1=2.2$ and $\omega=\sqrt5-2$ for the hopping model, and $V/t_1=7/8$ and $\omega=(\sqrt5-1)/4$ for the onsite model. Both parameter sets use $\delta=\pi\omega$. Derivations and additional numerical checks are given in the Supplemental Material (SM)~\cite{SMdeposit}, Secs.~\SIsection{sec:sm-growth}{1} and \SIsection{sec:sm-fig12}{4}.

For classification, the complex-phase dependence distinguishes subcritical and critical regimes that both have vanishing real-phase Lyapunov exponents (LEs). Under the phase continuation \(\theta_0\mapsto\theta_0+ih\), $L(E,h)$ measures exponential transfer growth at energy $E$ per two-site cell, with $h$ in radians. For this dimer chain, we obtain the three-branch form by combining transfer duality, determinant identities, and analytic cocycle theory~\cite{molinari2003,deninger2011,ge2026,ajs2014} (SM Sec.~\SIsection{sec:sm-growth}{1}). For $\lambda>0$ and real $E,h$, it reads
\begin{equation}
L(E,h)=\max\{A,\ B+|h|,\ C+2|h|\},\label{eq:three-branch}
\end{equation}
where $A=L(E,0)\geq0$ and $C(\lambda)=\log[\lambda^2/(4t_1t_2)]$. The affine branch $B+|h|$ has extrapolated $h=0$ intercept $B$, which can be negative; its active interval may have zero length. The zero-slope plateau has half-width $A-B\geq0$.
Fourier expansion maps the model to a dual chain of coupled momentum channels $\phi+2\pi n\omega$. Let $\gamma_1(E,\lambda)\geq\gamma_2(E,\lambda)$ denote its two largest Lyapunov exponents. The leading exponent then fixes $B=C+\gamma_1$; the transfer matrix and exact normalization are derived in SM Sec.~\SIsection{sec:sm-growth}{1}.

At a spectral energy $E$, Avila's classification~\cite{avila2015} identifies the physical transfer cocycle as subcritical when $A=0$ and $A-B>0$, critical when $A=B=0$, and supercritical when $A>0$. These regimes are illustrated by the LE profiles in Fig.~\ref{fig:abc-growth}(c). For the central-edge representatives in Fig.~\ref{fig:abc-growth}(b), finite-size scaling of the inverse participation ratio (IPR) provides complementary real-space evidence~\cite{hiramoto1992,evers2008}. Here $\mathrm{IPR}=\sum_j|\psi_j|^4$ for a normalized state on $N_s$ physical sites. The IPR of the recovered state scales as $N_s^{-1}$, whereas those of the neighboring localized controls approach nonzero values [Fig.~\ref{fig:abc-growth}(d)]. At the critical contact, the IPR decreases with system size and its finite-size exponents vary across the sampled sizes; the critical classification follows from the LE profile.

\emph{Spectral mechanism of reentrant localization.---}\phantomsection\label{sec:occupation}\label{sec:global-recovery}
On the bulk spectrum $\Sigma_\lambda$, the Jacobi spectral criterion excludes a positive LE with a plateau around $h=0$~\cite{ajs2014,avila2015,marx2014}. Equation~\eqref{eq:three-branch} therefore gives $A=\max(0,B)$ on the spectrum, and the subcritical population is exactly
\begin{equation}
 \Sigma_{{\rm sub},\lambda}
 =\Sigma_\lambda\cap\{E:B(E,\lambda)<0\}.
 \label{eq:spectral-occupation}
\end{equation}
Combined with $A=\max(0,B)$, Avila's classes reduce to $B<0$, $B=0$, and $B>0$ for subcritical, critical, and supercritical spectral energies, respectively.

The outer branch gives an upper modulation bound, $\lambda_H=2\sqrt{t_1t_2}$: $L(E,0)\geq C>0$ at every real energy for $\lambda>\lambda_H$. A fully supercritical spectrum at some $\lambda_a<\lambda_H$, followed by a subcritical spectral energy at $\lambda_b>\lambda_a$, therefore establishes global reentrance of a population with positive IDS weight. Continuity and full support of the density of states make the strict growth signs persist over modulation intervals. The global sufficient criterion is given in SM Sec.~\SIsection{sec:sm-limit}{1.3}; numerical checks of central-edge recovery are reported in SM Sec.~\SIsection{sec:sm-fig12}{4}.

Both the hopping-dimerized and onsite-staggered~\cite{chang2026} reference models realize this sequence at the central-gap edge. We track its upper edge $g(\lambda)$ against the selected energy branch $E=b(\lambda)$ of the zero-growth contour $B(E,\lambda)=0$ that meets the edge at capture and release [Fig.~\ref{fig:abc-growth}(b); SM Sec.~\SIsection{sec:sm-fig12}{4}].

\emph{Microscopic origin of capture and release.---}\phantomsection\label{sec:ordered-recovery}
In the hopping reference, Fourier scattering shifts the edge at fixed IDS label, whereas a neutral dual mode fixes the zero-growth branch $b(\lambda)$. The corresponding onsite auxiliary resonances and contacts are checked in SM Sec.~\SIsection{subsec:sm-onsite-resonance}{4.2}. Let $\mathcal N_\lambda(E)$ be the IDS per site~\cite{avronsimon1983} and $E_\lambda(N)=\inf\{E:\mathcal N_\lambda(E)\geq N\}$ its energy quantile. Spectral symmetry gives $g(\lambda)=E_\lambda(1/2+)$, with $1/2+$ the right-hand limit.

\begin{figure}[!t]
\centering
\includegraphics[width=\columnwidth,trim=0 .3mm 0 .2mm,clip]{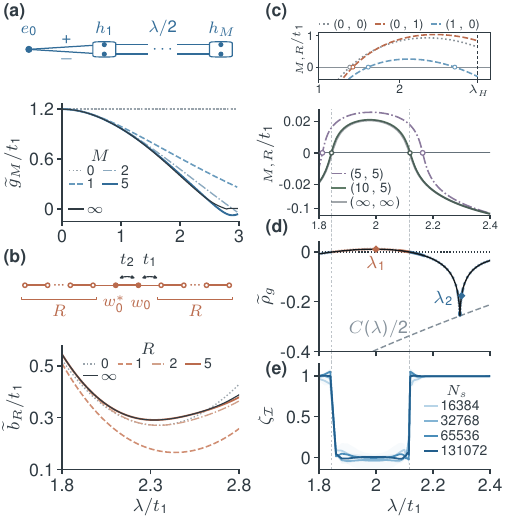}
\caption{Capture and release in the hopping model. (a) Edge estimates with $M$ physical Fourier channels per side. (b) Resonance estimates with $R$ auxiliary dimers per side; $\infty$: independent bulk references in (a,b). (c) Return mechanism: $\Delta_{M,R}=\widetilde g_M-\widetilde b_R$, with linear overview (upper) and asinh detail (lower). The first channel-return correction to $g$, $(M,R)=(1,0)$, already yields capture and release. Circles: refined contacts; upper dashed line: recovery ceiling $\lambda_H$. (d) Growth-rate check: mode estimate $\widetilde{\rho}_g$ (black), direct dual transfer (dots), and ordered controls (diamonds). Dotted: zero; dashed: channel-isolation threshold $C/2$. (e) Fixed-packet scaling $\zeta_{\mathcal I}$; lines/bands: two-phase mean/range. Gray guides in lower (c) and (d,e): bulk contacts. Methods: SM Secs.~\SIsection{subsec:sm-fig2}{4.1} and \SIsection{sec:sm-width-robustness}{4.4}.}
\label{fig:abc-shortchain}
\end{figure}

\phantomsection\label{sec:mechanism}
At $\lambda=0$, the upper central-gap edge has energy $e_0=t_2-t_1$, with state $u=(1,-1)^{\mathsf T}/\sqrt2$ at $k=\pi$. Finite modulation couples momenta $k_n=\pi+2\pi n\omega$, $n\in\mathbb Z$. Each channel carries an $A/B$ amplitude pair on which a $2\times2$ Hamiltonian block acts; neighboring channels couple with magnitude $\lambda/2$. Define the physical Fourier cutoff $M$ by retaining $-M\leq n\leq M$, giving a matrix of dimension $2(2M+1)$. Thus $M=0$ keeps only the bare edge channel. For $M\geq1$, scattering into the $M$ channels on each side and back produces a return self-energy. Combining the two tails by reflection symmetry gives
\begin{equation}
 \widetilde g_M=e_0+\frac{\lambda^2}{2}
 u^\dagger K_1(\widetilde g_M)^{-1}u ,
 \label{eq:edge-return-root}
\end{equation}
where $K_1(E)^{-1}$ is the Green matrix at the first tail channel, obtained by recursive block elimination through $n=1,\ldots,M$~\cite{sancho1985}. Following the root from $\widetilde g_0=e_0$ gives the estimates in Fig.~\ref{fig:abc-shortchain}(a). Independent variational spectral-distance estimates~\cite{davies2004spectral}, $\bar g_M\geq g$, decrease monotonically to $g$ and provide a reference for $\widetilde g_M$ (SM Sec.~\SIsection{sec:sm-returns}{2}).

For $b(\lambda)$, multiplying the dual transfer by $e^C$ shifts its leading growth to $B=C+\gamma_1$. At $B=0$, an isolated leading channel ($\gamma_1>\gamma_2$, equivalently $C+\gamma_2<0$) admits an analytic neutral mode. Fourier expansion of this neutral mode yields an auxiliary-chain eigenproblem whose selected resonance determines $b(\lambda)$ (SM Sec.~\SIsection{sec:sm-returns}{2}). The auxiliary sites are Fourier components of the dual mode. Their chain [Fig.~\ref{fig:abc-shortchain}(b)] has alternating hoppings $t_1,t_2$ and conjugate onsite terms about its center. The central pair, with onsite terms $w_0^*,w_0$, $w_0=\lambda\cos(\pi\omega/2+iC)$, and coupling $t_2$, gives the bare resonance $\widetilde b_0$. Define the auxiliary cutoff $R$ as the number of outer strong-bond dimers retained on each side: $R=0$ keeps this central pair, while $R>0$ adds return self-energies and gives $\widetilde b_R$ in a chain of $2+4R$ sites. Within the contact range, the central pair's decreasing mean energy and increasing real splitting lower the bare resonance and permit capture.

The resulting relative motion of the edge and resonance produces capture and release, measured by $\Delta_{M,R}=\widetilde g_M-\widetilde b_R$ [Fig.~\ref{fig:abc-shortchain}(c)]. Here $M$ counts physical momentum channels per side, whereas $R$ counts dimers in the dual-mode Fourier chain. Below $\lambda_H$, the bare construction $(M,R)=(0,0)$ has only one crossing. Including the nearest Fourier-channel returns, $0\to\pm1\to0$, produces a negative edge self-energy and lowers $\widetilde g_1$ back through $\widetilde b_0$, bringing release below $\lambda_H$ already at $(1,0)$. By contrast, adding only the first outer auxiliary dimers, $(0,1)$, retains a single crossing below this ceiling; these complete crossing sequences are shown in the SM. Further returns at $(5,5)$ and $(10,5)$ bring both contacts toward the independent bulk values. The minimal construction explains the crossing sequence; the deeper curves remain finite-return approximations. Recursions, contact values, and onsite-model checks are given in SM Secs.~\SIsection{sec:sm-returns}{2} and \SIsection{sec:sm-fig12}{4}.

The crossing mechanism predicts capture and release at $g=b$, where the edge growth changes sign. Figure~\ref{fig:abc-shortchain}(d) tests this connection numerically through
\begin{equation}
 \rho_g(\lambda)=B(g(\lambda),\lambda)
 =C(\lambda)+\gamma_1(g(\lambda),\lambda).
 \label{eq:edge-rate}
\end{equation}
Wherever the mode branch is resolved, the mode estimate $\widetilde{\rho}_g$ agrees with direct dual-transfer products and reproduces the sign changes across the bulk contacts. Independent calculations locate capture and release near $\lambda/t_1=1.85,2.12$ [guides in Fig.~\ref{fig:abc-shortchain}(c--e)]. Positive/negative $\rho_g$ identifies supercritical/subcritical edge states; a later return to $\rho_g>0$ marks recapture. At $\lambda/t_1=2.00,2.30$, $\rho_g\simeq0.0116,-0.1758$, respectively, confirming a supercritical edge followed by recovery at the two ordered controls.

More generally, for fixed $\omega$, a positive-weight interval of IDS labels immediately above $N=1/2$ recovers if $0<\lambda_1<\lambda_2<\lambda_H$ and
\begin{equation}
 \rho_g(\lambda_1)>0,\qquad
 \frac{C(\lambda_2)}{2}<\rho_g(\lambda_2)<0.
 \label{eq:edge-two-point}
\end{equation}
The lower bound keeps the leading dual mode isolated; the negative sign identifies recovery. Continuity extends the criterion to a common interval of IDS labels $\mathcal I=(1/2,1/2+\varepsilon)$, $\varepsilon>0$, specifying the same spectral packet at both controls.

To test this bulk prediction in finite chains, Fig.~\ref{fig:abc-shortchain}(e) follows participation scaling in the fixed IDS interval $\mathcal I=[0.50025,0.50075]$ as $N_s$ increases. All states with IDS-midpoint labels in this interval are retained at each size, so the number of sampled states grows with $N_s$. The exponent $\zeta_{\mathcal I}=\log_2[\langle P_{2N_s}\rangle_{\mathcal I}/\langle P_{N_s}\rangle_{\mathcal I}]$ compares successive size doublings, where $P=1/\mathrm{IPR}$ and $\langle\cdot\rangle_{\mathcal I}$ is the packet average at fixed phase. At the two ordered controls, $\zeta_{\mathcal I}$ stays near zero and one over the tested sizes, respectively, consistent with localized and recovered extended states. Wider-packet tests show the same behavior in both models (SM Sec.~\SIsection{sec:sm-width-robustness}{4.4}).

\begin{figure}[!t]
\centering
\includegraphics[width=\columnwidth,trim=0 .6mm 0 .25mm,clip]{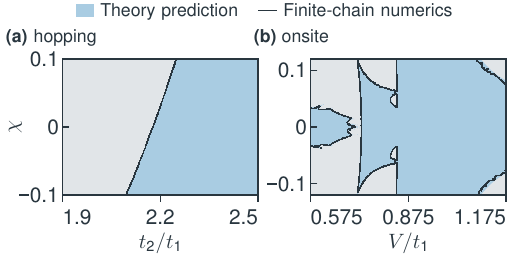}
\caption{Regions of reentrant recovery in (a) hopping and (b) onsite models. Blue: theory prediction from Eq.~\eqref{eq:edge-two-point}; dark lines: numerical recovery boundaries from a separate real-space diagonalization and participation protocol for the fixed IDS packet; gray: no ordered pair satisfying Eq.~\eqref{eq:edge-two-point} detected by the stated search. Here $\chi=\delta-\pi\omega$. Participation criteria and finite-size checks: SM Sec.~\SIsection{sec:sm-fig12}{4}.}
\label{fig:parameter-test}
\end{figure}

\phantomsection\label{sec:parameter-tests}
Reentrant recovery persists over finite parameter ranges for both gap-opening mechanisms, including nonzero relative-phase detuning [Fig.~\ref{fig:parameter-test}]. The maps vary (a) hopping dimerization $r=t_2/t_1$ at $V=0$ and (b) onsite staggering $V/t_1$ at $r=1$, each against $\chi=\delta-\pi\omega$. The regions predicted by Eq.~\eqref{eq:edge-two-point} (blue) have boundaries close to finite-chain participation boundaries (dark) for the prescribed packet $\mathcal I=[0.50025,0.50075]$. On the shared grid, separate real-space diagonalization and participation analysis detect recovery at $98.1\%$ and $96.7\%$ of coordinates satisfying the growth criterion in the hopping and onsite models, respectively. These conditional rates depend on the specified packet, chain sizes, control searches, and participation thresholds. Some discrepancies remain unresolved (SM Sec.~\SIsection{subsec:sm-parameter-planes}{4.3}).

\setcounter{topnumber}{1}

\emph{Exact control of spectral crossings.---}\phantomsection\label{sec:prediction}
The recovery window can be tuned by pinning recapture and shifting release through the relative motion of the spectral edge $g$ and the zero-growth boundary $b$. In the onsite-staggered reference, compensated sublattice modulation implements this control, transporting the selected contacts with their IDS labels and zero-growth conditions intact. The hopping counterpart, including window control and IDS-resolved thresholds, is given in SM Sec.~\SIsection{subsec:sm-control-checks}{4.5}. Specifically, we follow the central edge at $N_*=1/2^+$, using $V$ for compensation and sublattice amplitudes $\lambda e^{\pm\eta}$:
\begin{equation}
\begin{aligned}
u_{A,m}&=-V+\lambda e^\eta\cos(\theta_m-\pi\omega+\psi),\\
u_{B,m}&= V+\lambda e^{-\eta}\cos\theta_m,
\end{aligned}
\label{eq:control-model}
\end{equation}
where $\eta$ controls the amplitude contrast and $\eta=0$ recovers the equal-amplitude reference. The fixed amplitude product $\lambda^2$ permits an exact pointwise map to the equal-amplitude chain with the same $A$--$B$ hoppings. The map holds at fixed $\psi$; we set $\psi=0$, hence $\delta=\pi\omega$, in the reference example.

\begin{figure}[!t]
\centering
\includegraphics[width=\columnwidth,trim=0 .2mm 0 1.2mm,clip]{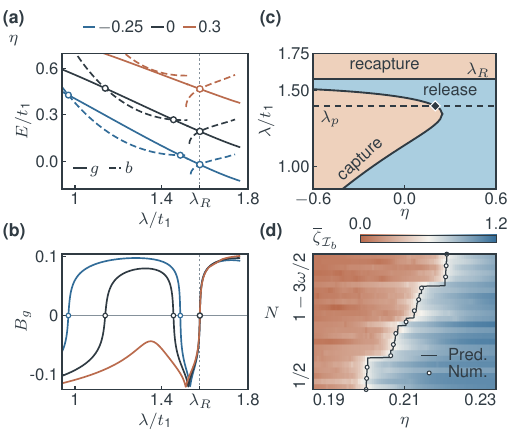}
\caption{Compensated recovery in the onsite model. (a) Edge $g$ (solid) and neutral boundaries $b$ (dashed) for $\eta=-0.25,0,0.30$. Circles: contacts; vertical guide: pinned $\lambda_R$. (b) Corresponding edge growth $B_g=B_\eta(g,\lambda)$ on an asinh scale. At $\eta=0.30$, only recapture remains. (c) Blue/orange: $B_g<0$/$B_g>0$. Dashed: $\lambda_p=1.40t_1$; diamond: release intersection. (d) Four-phase mean participation exponents $\overline\zeta_{\mathcal I_b}$ for $8362\to16724$ sites versus $\eta$ and IDS label $N$. Line/circles: predicted/direct target thresholds. Here $\mathcal I=(1/2,1-3\omega/2)$, $\omega=(\sqrt5-1)/4$. Methods and endpoint-side tests: SM Secs.~\SIsection{sec:sm-control}{3}, \SIsection{sec:sm-target-validation}{4.5}.}
\label{fig:abc-prediction}
\end{figure}

\begingroup
\predisplaypenalty=0
Let $H_\eta(V)$ denote the chain with potentials~\eqref{eq:control-model}. With $D_\eta=\operatorname{diag}(e^{-\eta/2},e^{\eta/2})$, the rescaling gives:
\begin{equation}
\begin{gathered}
D_\eta[H_\eta(V)-E]D_\eta=H_0(V')-E',\\
\begin{pmatrix}E'\\V'\end{pmatrix}
=\begin{pmatrix}
\cosh\eta&-\sinh\eta\\
-\sinh\eta&\cosh\eta
\end{pmatrix}
\begin{pmatrix}E\\V\end{pmatrix}.
\end{gathered}
\label{eq:control-map}
\end{equation}
\endgroup
Here $H_0(V')$ is the equal-amplitude source, evaluated separately at each test energy because $V'$ depends on $E$. The map is a hyperbolic rotation in $(E,V)$, preserving $E^2-V^2$. At matched parameters, the congruence preserves spectral membership and IDS, and transfer-matrix similarity preserves $L(E,h)$ and $B$. Together, these relations transport the spectral contacts underlying the crossing mechanism from source to target. In particular, fixing a source recapture contact $(E_c,V_c,\lambda_R)$ on branch $N_*$ defines the compensation path
\begin{equation}
\begin{aligned}
V(\eta)&=V_c\cosh\eta+E_c\sinh\eta,\\
E_R(\eta)&=E_c\cosh\eta+V_c\sinh\eta .
\end{aligned}
\label{eq:control-anchor}
\end{equation}
Along this path, the central edge $g_\eta$ and the recapture branch $b_{\eta,R}$ satisfy
\begin{equation}
\begin{gathered}
g_\eta(\lambda_R)=b_{\eta,R}(\lambda_R)=E_R(\eta),\\
B_\eta(E_R(\eta),\lambda_R)=0.
\end{gathered}
\label{eq:control-pinned-crossing}
\end{equation}
The contact therefore stays at $\lambda_R$ while it remains regular, with its energy moving along $E_R(\eta)$ [Fig.~\ref{fig:abc-prediction}(a,c)].

Control of release follows from the differential motion of $g$ and $b$ along this compensation path. For the controls in Fig.~\ref{fig:abc-prediction}(a), increasing $\eta$ raises both energies on the capture--release branch, with $b$ rising more strongly. The positive lobe of $g-b$ shrinks, delaying capture and advancing release, while recapture stays pinned. To calculate the moving release endpoint $\lambda_L$ at fixed IDS, let $e_N(\lambda,v)$ be the source energy of label $N$. The inverse map uniquely determines the matched source staggering $v_N$ from
\begin{equation}
 V(\eta)=v_N\cosh\eta+e_N(\lambda,v_N)\sinh\eta .
\label{eq:control-source-transport}
\end{equation}
At $N=N_*$, growth invariance locates release by transporting its source $B=0$ contact. At a simple contact, $d\lambda_j/d\eta=-\partial_\eta(g-b)/\partial_\lambda(g-b)$, with $b$ the corresponding branch satisfying $B_\eta(b,\lambda)=0$ and the compensation included in the $\eta$ derivative (SM Sec.~\SIsection{sec:sm-control}{3}). For the width $W=\lambda_R-\lambda_L$, the source has $W(0)=0.1220t_1$. The full map gives $57.3\%$ widening at $\eta=0.225$ and $26.0\%$ narrowing at $\eta=-0.25$; direct contact root finding gives the curves in Fig.~\ref{fig:abc-prediction}(c). Near $\eta=0.247$, capture and release merge as the positive lobe disappears. Beyond this merger, the recovered band connects to the weak-modulation subcritical regime, with $\lambda_R$ still marking recapture.

The same zero-growth condition predicts recovery for other IDS labels, with the selected spectral energy taking the place of the central edge. At fixed probe modulation $\lambda_p$, we follow labels in $\mathcal I=(1/2,1-3\omega/2)$, between the central and next principal gaps. Write $B_0(E,\lambda,v)$ for the source growth intercept. For each label $N$, find a source contact $v_*(N)$ satisfying $B_0[e_N(\lambda_p,v_*),\lambda_p,v_*]=0$. Equations~\eqref{eq:control-anchor} and \eqref{eq:control-source-transport} then give the recovery threshold $\eta_*(N)$:
\begin{equation}
\tanh\eta_*(N)=
\frac{V_c-v_*(N)}
{e_N(\lambda_p,v_*(N))-E_c}.
\label{eq:control-population}
\end{equation}
A finite threshold exists when this ratio is defined and lies in $(-1,1)$. If each label has one positive-to-negative crossing as $\eta$ increases, $f_{\mathcal I}(\eta)=|\mathcal I|^{-1}\int_{\mathcal I}\mathbf1_{\{\eta>\eta_*(N)\}}\,dN$. For example, at $\lambda_p=1.40$, all 305 packet labels of the 8362-site periodic approximant cross once over $-0.6\leq\eta\leq0.6$. The threshold distribution spans $\Delta\eta\simeq0.018$; three sampled recovered fractions are reported in SM Sec.~\SIsection{sec:sm-target-validation}{4.5}.

Independent real-space diagonalization tests the predicted recovery order and window control. At $\lambda_p=1.40$, we retain all 305 labels of the unequal-amplitude target and scan 81 contrasts with $\Delta\eta=0.0005$. Fixed IDS groups recover in the predicted order [colors in Fig.~\ref{fig:abc-prediction}(d)]. For each group, we use the same participation exponent $\zeta_{\mathcal I}$ as in Fig.~\ref{fig:abc-shortchain}(e). The overlaid source thresholds and 13 directly located target contacts provide the growth-based reference. Direct target diagonalization also tests both sides of release and recapture at $\eta=-0.25,0,0.225$. For the fixed narrow packet $\mathcal I_e=[0.50025,0.50075]$, the corresponding $\zeta_{\mathcal I_e}$ changes from localized to extended scaling across release and reverses across recapture (SM Sec.~\SIsection{sec:sm-target-validation}{4.5}). These tests use offsets $\pm0.02t_1$ from the predicted endpoints. Full-packet size scaling and endpoint checks up to $262144$ sites are given in SM Sec.~\SIsection{sec:sm-target-validation}{4.5}.

\emph{Conclusion.---}Spectral-edge motion relative to a neutral auxiliary resonance explains capture and release of a common IDS packet in quasiperiodic dimer chains. Compensated sublattice modulation controls the recovery window by pinning recapture and shifting release. The exact sublattice mapping also predicts recovery thresholds for individual IDS labels from the zero-growth condition. Direct real-space scaling of the target Hamiltonian tests the predicted recovery order and endpoint control.

\begin{acknowledgments}
\textit{Acknowledgments.---}This work was supported by the National Natural Science Foundation of China under Grants No. 11974205, No. 61727801, and No. 62131002, and by the Key Research and Development Program of Guangdong Province (2018B030325002).
\end{acknowledgments}
\paragraph*{Data availability.} The data supporting this study are available from the authors upon reasonable request.
\bibliographystyle{apsrev4-2}
\AddToHook{cmd/thebibliography/after}{\rightskip=0pt plus 1em\relax}
\bibliography{references}
\end{document}